\documentclass[aps,prb,twocolumn,footinbib,superscriptaddress]{revtex4-2}
\usepackage{xcolor,soul}
\usepackage{graphicx}
\usepackage{dcolumn}
\usepackage{bm}
\usepackage[colorlinks=true,
            linkcolor=blue,
	        citecolor=blue,
	         urlcolor=blue]{hyperref}
\usepackage{mwe}  
\usepackage{chngcntr}  
\usepackage[mathlines]{lineno}
\usepackage[english]{babel} 
\usepackage[T1]{fontenc}
\usepackage{gensymb}
\usepackage{braket}
\usepackage[caption=false]{subfig}
\usepackage[version=4,arrows=pgf-filled,
]{mhchem}
\usepackage{siunitx}
\DeclareSIUnit{\rydberg}{Ry}
\DeclareSIUnit{\angstrom}{\textup{\AA}}
\DeclareSIUnit{\bar}{bar}
\usepackage[capitalise]{cleveref}  
\usepackage{svg}  
\usepackage{multirow}
\usepackage{tabularx}

\newcommand{\vk}{\boldsymbol{k}}

\newcommand{\vE}{\boldsymbol{E}}
\newcommand{\vq}{\boldsymbol{q}}

\newcommand{\vv}{\boldsymbol{v}}
\newcommand{\Fph}{F^{\text{ph}}}
\newcommand{\Fel}{F^{\text{el}}}
\newcommand{\Felex}{F^{\text{el}, *}}
\newcommand{\epcg}{g_{mn\nu}(\vk,\vq)}

\newcommand{\vectorr}{\boldsymbol{r}}

\newcommand{\abinitio}{\textit{ab initio}}

\newcommand{\kB}{k_{\mathrm{B}}}  
\newcommand{\CV}{C_{\text{V}}}  
\newcommand{\CP}{C_{\text{P}}}
\newcommand{\CVph}{\CV^{\text{ph}}}
\newcommand{\CVel}{\CV^{\text{el}}}
\newcommand{\BA}{B_{\text{A}}}
\newcommand{\BAprime}{\BA'}

\newcommand{\Vdft}{V^{\text{DFT}}}

\addto\captionsenglish{}
\crefname{section}{Sec.}{Secs.}

\begin{document}

\preprint{APS/123-QED}

\title{Effects of Spin-Orbit Coupling and Thermal Expansion on the Phonon-limited Resistivity of Pb from First Principles}%

\author{Félix Antoine Goudreault}
\affiliation{Département de Physique et Institut Courtois, Université de Montréal, C. P. 6128, Succursale Centre-Ville, Montréal, Québec, H3C 3J7, Canada}
\author{Samuel Poncé}
\affiliation{European Theoretical Spectroscopy Facility, Institute of Condensed Matter and Nanosciences (IMCN), Université catholique de Louvain (UCLouvain), 1348 Louvain-la-Neuve, Belgium}
\affiliation{WEL Research Institute, Avenue Pasteur 6, 1300 Wavre, Belgium}
\author{Feliciano Giustino}
\affiliation{Oden Institute for Computational Engineering and Sciences, The University of Texas at Austin, Austin, Texas 78712, USA}
\affiliation{Department of Physics, The University of Texas at Austin, Austin, Texas 78712, USA}
\author{Michel Côté}
\affiliation{Département de Physique et Institut Courtois, Université de Montréal, C. P. 6128, Succursale Centre-Ville, Montréal, Québec, H3C 3J7, Canada}

\date{\today}

\begin{abstract}
    Using density functional theory calculations with spin-orbit coupling (SOC), we report on the temperature-dependent thermodynamical properties of Pb: electrical resistivity, thermal expansion (TE), heat capacity, bulk modulus and its pressure derivative. 
    For the former, we employed the state-of-the-art \abinitio{} Boltzmann Transport Equation formalism, and we calculated the effect of TE. In accordance with previous work, we show that SOC improves the description of the phonon dispersion and the resistivity. 
    We argue that this is caused by a joint mutual effect of an increase in the electronic nesting and an increase in the electron-phonon coupling. 
    Interestingly, including TE incorporates non-linearity into the resistivity at high temperatures, whose magnitude depends on whether SOC is included or not. 
    We suggest that mechanisms beyond the quasi-harmonic approximation should be considered to get a better description of Pb with SOC at high temperatures.
\end{abstract}

\maketitle

\section{Introduction}
\label{sec:intro}

The thermodynamic properties of bulk face center cubic lead, Pb, have been studied in detail over the past century. 
Indeed, as one of the simplest elemental metals with a single atom in its unit cell, Pb is a well-known conventional superconductor with one of the highest transition temperature and electron-phonon coupling (EPC) strength~\cite{schrieffer_effective_1963,scalapino_strong-coupling_1966,savrasov_electron-phonon_1996,de_gironcoli_lattice_1995,floris_two-band_2007}.

Furthermore, Pb exhibits strong spin-orbit coupling (SOC) effects in both its electronic and vibrational degrees of freedom~\cite{verstraete_density_2008,heid_effect_2010,corso_ab_2008,vosko_influence_1965,liu_linear-response_1996,smirnov_effect_2018,ponce_epw:_2016,sklyadneva_electron-phonon_2012}. 
As such, from the point of view of \abinitio\ calculations, Pb was a prime candidate to be studied within the framework of Density Functional Perturbation Theory (DFPT) in its early years~\cite{savrasov_electron-phonon_1996,de_gironcoli_lattice_1995,liu_linear-response_1996} and in the past decade when SOC was added to the DFPT frameworks~\cite{corso_ab_2008,verstraete_density_2008,heid_effect_2010,sklyadneva_electron-phonon_2012}. 
It was shown that SOC is needed to correctly describe the finer structures in the phonon dispersion, such as Kohn anomalies~\cite{corso_ab_2008,verstraete_density_2008,heid_effect_2010,sklyadneva_electron-phonon_2012,smirnov_effect_2018}, where a general phonon softening throughout the whole first Brillouin Zone, particularly pronounced at the $\vq=X$ point, was observed. 
The overall results agree better with inelastic neutron scattering data taken at \qty{100}{\kelvin} than with the results obtained without SOC~\cite{brockhouse_crystal_1962,brockhouse_old_1961}. 
Generally, SOC also changes the EPC matrix elements which induces striking changes in the Eliashberg spectral function and the electronic self-energy~\cite{heid_effect_2010,sklyadneva_electron-phonon_2012}.

Over the years, many thermodynamical properties of Pb have been computed from first principles, including free energy~\cite{grabowski_ab_2007,smirnov_effect_2018}, equation of states~\cite{strassle_equation_2014,smirnov_effect_2018}, phonon dispersion and EPC strength~\cite{savrasov_electron-phonon_1996,de_gironcoli_lattice_1995,liu_linear-response_1996,corso_ab_2008,heid_effect_2010,verstraete_density_2008,grabowski_ab_2007,ponce_epw:_2016,rittweger_phonon_2017,sklyadneva_electron-phonon_2012}, thermal expansion (TE)~\cite{grabowski_ab_2007} and transport properties~\cite{savrasov_electron-phonon_1996,ponce_epw:_2016,smirnov_effect_2018,rittweger_phonon_2017}. 
Regarding the latter, the published calculated phonon-limited temperature-dependent resistivity curves, albeit close to experiments, all present some discrepancies at high temperatures, which remains an open question. 
%
%
Instead, for harmonic vibrational properties, most of the disagreements with respect to experimental data seen in the earlier work of \citet{savrasov_electron-phonon_1996} were accounted for by including SOC as shown by \citet{smirnov_effect_2018} and \citet{ponce_epw:_2016}. 

Additionally, the reported \abinitio\ computations of lead's resistivity were all done within approximated solutions to the Boltzmann Transport Equation (BTE), namely the Lowest-Order Variational Approximation (LOVA)~\cite{allen_new_1978,ziman_electrons_1960} used in Refs.~\cite{savrasov_electron-phonon_1996,ponce_epw:_2016,smirnov_effect_2018} or the Momentum Relaxation Time Approximation (MRTA) and the constant Relaxation Time Approximation (cRTA) used in Ref.~\cite{rittweger_phonon_2017}. 
As all these methods are different approximations of a more general approach called the \abinitio{} BTE (aiBTE)~\cite{ponce_first-principles_2020} and because a recent report showed significant differences between the aiBTE, MRTA, and cRTA schemes~\cite{claes_assessing_2022}, we have yet to see if a higher level of theory could explain these discrepancies observed in Pb at high temperatures.

In this work, we use Density Functional Theory (DFT) within the Quasi Harmonic Approximation (QHA) to compute the effect of TE and SOC on \ce{Pb}'s thermodynamical properties, namely the bulk modulus and its pressure derivative, the heat capacity and the phonon-limited temperature dependence of its electrical resistivity. 
The latter is computed using the state-of-the-art aiBTE formalism which has not been applied to \ce{Pb} so far. 
Furthermore, to the authors' knowledge, we first report the computation of \ce{Pb}'s TE, heat capacity, and bulk modulus pressure derivative with SOC. 
The theoretical framework is laid out in Sec.~\ref{sec:method} and the computational methodology in Sec.~\ref{sec:computational_details}. The results obtained are shown and discussed in Sec.~\ref{sec:results}.
Finally, we present our conclusions and outlooks.

\section{Method}
\label{sec:method}

\subsection{Thermodynamic properties}

The thermodynamic properties of Pb can be computed from the Helmholtz free energy $F(V, T)$ curves of the system where $V$ is the unit cell volume and $T$ the temperature. 
In a system comprised of non-interacting electrons and phonons, the free energy can be decomposed into electronic $\Fel$ and phononic $\Fph$ terms, which yields~\cite{ponce_hole_2019,palumbo_lattice_2017,baroni_density-functional_2010}:
\begin{equation}
    \label{eq:free_energy}
    F(V,T)=F^{\text{ph}}(V,T) + F^{\text{el}}(V,T),
\end{equation}
where $F^{\text{el}}$ can further be decomposed in two contributions: $F^{\text{el}}(V,T) = U(V) + F^{\text{el},*}(V, T)$ where $U(V)$ is the clamped-ion energy at~\qty{0}{\kelvin} and $F^{\text{el},*}(V,T)$ is the free energy due to thermal electronic excitations~\cite{ponce_hole_2019}. 
The $F^{\text{el},*}(V,T)$ term is typically neglected in insulators when the electronic energy gap is much larger than thermal energies.
In metals, $F^{\text{el},*}$ cannot be discarded and we rely on the finite-temperature formulation of DFT introduced by \citet{mermin_thermal_1965}, which was reported to be the most accurate method for computing  $F^{\text{el},*}$ compared to other approximations such as the density of states (DOS) approach or the Sommerfeld approximation~\cite{zhang_accurate_2017}. 
As such, for each sampled unit cell volume and temperature, we set $F^{\text{el}}(V, T)$ as the total energy of a ground-state DFT calculation where the single-particle occupation numbers are determined by a Fermi-Dirac distribution.

The vibrational contribution to the Helmholtz free energy can be written within the QHA as~\cite{palumbo_lattice_2017,ponce_hole_2019}: 
\begin{align}\label{eq:free_energy_phonons}
  F^{\rm ph}(V,T) = \frac{k_{\rm B}T}{N_q}\sum_{\substack{\vq\nu\\\omega_{\vq\nu}(V)\neq 0}}\ln\left[2\sinh\left(\frac{\hbar\omega_{\vq\nu}(V)}{2k_{\rm B}T}\right)\right],
\end{align} 
where $N_q$ is the number of $\vq$ points in the first Brillouin Zone, $\omega_{\vq\nu}(V)$ is the phonon frequency for branch $\nu$ and wavevector $\vq$ for a given unit cell volume $V$ and $\hbar$ and $k_{\rm B}$ are respectively the reduced Planck's and Boltzmann's constants.

Once $F(V, T)$ is known for a given set of temperatures and volumes, we extract the various thermodynamical properties by fitting it to the third-order Birch-Murnaghan isothermal equation of state which can be written as~\cite{birch_finite_1947,murnaghan_compressibility_1944}:
\begin{multline} \label{eq:birch-murnaghan}
        F(V, T) = F_0(T) + \frac{9 B(T) V_0(T)}{16}\left[\left(\frac{V_0(T)}{V}\right)^{\frac{2}{3}}-1\right]^2\\
        \times\left\{\left[\left(\frac{V_0(T)}{V}\right)^{\frac{2}{3}}-1\right]B'(T) + \left[6-4\left(\frac{V_0(T)}{V}\right)^{\frac{2}{3}}\right]\right\},
\end{multline}
where $F_0$, $V_0$, $B$ and $B'$ are the minimum of the free energy, the equilibrium volume, the isothermal bulk modulus and the pressure derivative, respectively. 
By extracting $V_0(T)$ from the fit at each sampled temperature, we obtain the temperature-dependent lattice parameter $a(T)$. 

The heat capacity at constant volume can be derived from the free energy using $C_{\rm V}=-T \frac{\partial^2F}{\partial T^2}\big|_{\rm V}$. 
As $F$ consists of a phonon term and an electronic term, we can also compute $C_{\rm V}(T)=C_{\rm V}^{\text{ph}}(T) + C_{\rm V}^{\text{el}}(T)$ separately. 
For phonons, we have
\begin{equation}
    \label{eq:cv_phonons}
    C_{\rm V}^{\text{ph}}(T)= k_{\rm B}\sum_{\vq\nu}\left(\frac{\hbar\omega_{\vq\nu}}{2k_{\rm B}T}\right)^2\sinh^{-2}\left(\frac{\hbar\omega_{\vq\nu}}{2k_{\rm B}T}\right),
\end{equation}
while the electronic part $C_V^{\text{el}}$ is computed using a finite difference formula and interpolated at the equilibrium volume $V_0(T)$. 
Most experiments are instead made at constant pressure which can be obtained as~\cite{wallace_thermodynamics_1998}:
\begin{equation}
    \label{eq:heat_capacity_constant_pressure}
    C_{\rm P}(T) = C_{\rm V}(T) + \alpha^2B V_0(T) T,
\end{equation}
where $B$ is taken from the free energy fits of \cref{eq:birch-murnaghan} and $\alpha(T)=\frac{1}{V_0} \frac{\partial V_0(T)}{\partial T}\big|_{\rm P}$ is the linear TE coefficient, evaluated at constant pressure $P$.

\Cref{eq:birch-murnaghan} allows the computation of the isothermal bulk modulus and its pressure derivative. 
But, as for the heat capacity, what is typically measured experimentally is the adiabatic bulk modulus, $B_{\rm A}$, and its pressure derivative, $B_{\rm A}'$. 
The relationships between $B_{\rm A}$ and $B$ are~\cite{wallace_thermodynamics_1998}

\begin{align}
    \label{eq:theory_adiabatic_bulk_modulus}
    \BA =& B \frac{\CP}{\CV}\\
    \label{eq:theory_adiabatic_bulk_modulus_derivative}
    \BAprime =& \left(\frac{\BA}{B}\right)^2B' + \BA^2  \left.\frac{\partial\Delta k}{\partial P}\right|_T,
\end{align}
\noindent
where $\Delta k=1/B - 1/B_{\rm A}$ and 


\begin{equation}
\label{eq:theory_adiabatic_bulk_modulus_derivative_offset}
        \frac{\partial\Delta k}{\partial P}\bigg|_T = \Delta k\biggl\{\Delta k \left[1 +\frac{1}{\alpha^2} \left.\frac{\partial \alpha}{\partial T}\right|_{\text{P}}\right]
        +\frac{2}{\alpha B^2} \left.\frac{\partial B}{\partial T}\right|_{\rm P}-\frac{1}{B}\biggr\}.
\end{equation}

\subsection{Electronic Transport}
\label{sec:theory_electronic_transport}

The phonon-limited DC conductivity can be obtained by solving the full aiBTE as~\cite{ponce_first-principles_2020}
\begin{equation}
    \label{eq:conductivity_tensor}
    \sigma_{\alpha\beta} = -\frac{e}{(2\pi)^3}\sum_n\int d^3\vk v_{n\vk,\alpha}\partial_{E_\beta}f_{n\vk},
\end{equation}
\noindent
where $e$ is the electron charge, $\partial_{E_\beta}f_{n\vk}$ is the derivative of the occupation factor for the electronic state $\ket{n\vk}$, where $n$ is the band index and $\vk$ the wavevector, with respect to the electric field $\vE$ and $v_{n\vk,\alpha}=\braket{n\vk|(i/\hbar)[\hat{H}, r_\alpha]|n\vk}$ is the diagonal element of the electronic band velocity operator for the state $\ket{n\vk}$ with $\hat{H}$ being the Kohn-Sham Hamiltonian and $\vectorr$ being the position operator~\cite{starace_length_1971}. 
Considering the latter, in this work, we explicitly compute the commutator in order to correctly capture the effects of the non-local part of the potential~\cite{read_calculation_1991,ismail-beigi_coupling_2001,pickard_nonlocal_2003,ponce_epw:_2016,ponce_first-principles_2021_hall}. 
%
%
Without external fields, the occupation factor $f_{n\vk}$ reduces to the Fermi-Dirac distribution denoted as $f_{n\vk}^0$. 
In \cref{eq:conductivity_tensor}, the integral is carried out over the first Brillouin Zone, Greek indices denote Cartesian directions, and the temperature dependence is contained within the occupation factor term. 
The $\partial_{E_\beta}f_{n\vk}$ term can be computed using the linearized BTE formalism where, in the context of EPC scattering without magnetic fields and temperature gradients, it can be computed iteratively from~\cite{ponce_towards_2018,ponce_first-principles_2020}:
\begin{multline}\label{eq:ibte}
        \tau_{n\vk}^{-1}\partial_{E_\beta} f_{n\vk} = e\frac{\partial f_{n\vk}^0}{\partial\epsilon_{n\vk}}v_{n\vk,\beta}+\frac{2\pi}{\hbar}\sum_{m\nu}\int\frac{d^3\vq}{\Omega_{\text{BZ}}}|g_{mn\nu}(\vk,\vq)|^2\\
        \times[(1+n_{\vq\nu}-f^0_{n\vk})\delta(\Delta_{\vk,\vq}^{nm}+\hbar\omega_{\vq\nu})\\
 +(n_{\vq\nu}+f^0_{n\vk})\delta(\Delta_{\vk,\vq}^{nm}-\hbar\omega_{\vq\nu})]\partial_{E_\beta}f_{m\vk+\vq},  
\end{multline}
where $\Omega_{\text{BZ}}$ is the Brillouin Zone volume, $g_{mn\nu}(\vk,\vq)$ is the EPC matrix element which measures the coupling strength between states $\ket{n\vk}$ and $\ket{m\vk+\vq}$ via a phonon with a wavevector $\vq$ on branch index $\nu$, $n_{\vq\nu}$ is the Bose-Einstein occupation factor for that phonon mode and $\Delta_{\vk,\vq}^{nm}=\epsilon_{n\vk}-\epsilon_{m\vk+\vq}$ is the single-particle electron eigenvalue difference between the initial and scattered states which are obtained from DFT. 
The scattering rate $\tau_{n\vk}^{-1}$ is defined as
\begin{multline} \label{eq:relaxation_time}
        \tau_{n\vk}^{-1} = \frac{2\pi}{\hbar}\sum_{m\nu}\int\frac{d^3\vq}{\Omega_{\text{BZ}}}|g_{mn\nu}(\vk,\vq)|^2\\
        \times[(1-f^0_{m\vk+\vq}+n_{\vq\nu})\delta(\Delta_{\vk,\vq}^{nm}-\hbar\omega_{\vq\nu})\\
         +(f^0_{m\vk+\vq}+n_{\vq\nu})\delta(\Delta_{\vk,\vq}^{nm}+\hbar\omega_{\vq\nu})].
\end{multline}

In \cref{eq:ibte,eq:relaxation_time}, the EPC matrix element is computed at the linear-response level using the perturbed Kohn-Sham potential $V$ as $\epcg~=~\braket{m\vk+\vq|\Delta_{\vq\nu}V|n\vk}$~\cite{giustino_electron-phonon_2017}. 
This formalism is at a higher level of theory than the so-called LOVA approach~\cite{ponce_first-principles_2020}. 
Thus, it should be more accurate as the latter may be derived by averaging out the relaxation times on the Fermi surface and assuming all electrons within the partially filled conduction band participate equally to the transport~\cite{ziman_electrons_1960,grimvall_electron-phonon_1981,allen_new_1978,ponce_first-principles_2020} which might be inaccurate for materials with complex anisotropic Fermi surfaces such as Pb.

\subsection{Fermi Surface Nesting and Phonon Linewidth}
\label{subsec:fs_nesting_ph_linewidth}

The inclusion of SOC in the phonon calculation of Pb leads to phonon softening due to an increased electronic nesting within the Fermi surface~\cite{verstraete_density_2008,ponce_epw:_2016,heid_effect_2010,corso_ab_2008}, which can be computed using
\begin{equation}
    \label{eq:nesting_function}
    \zeta_{\vq} = s\sum_{nm}\zeta_{nm\vq}\equiv \!\frac{s}{N_k}\! \sum_{nm\vk}\!\delta(\epsilon_{n\vk}-\epsilon_{\rm F})\delta(\epsilon_{m\vk+\vq}-\epsilon_{\rm F}),
\end{equation}
where $\epsilon_{\rm F}$ is the Fermi energy, $N_k$ is the number of $\vk$-points in the Brillouin Zone and $s$ is a spin factor, which is equal to \num{4} without SOC and \num{1} with it. 
We also define the band-resolved Fermi surface nesting $\zeta_{nm\vq}$ where $n$ and $m$ are band indices. 
A drawback of the nesting function is that it does not take into account the EPC matrix elements. 
Indeed, the nesting function acts like a joint density of states, and in order to get a more accurate picture one can compute the phonon linewidths which, at low temperature, can be approximated as~\cite{allen_neutron_1972,giustino_electron-phonon_2017}:

\begin{multline} \label{eq:phonon_linewidth}
    \gamma_{\vq\nu}  = \frac{2\pi \omega_{\vq\nu}}{N_{\vk}}  \\
   \times \sum_{\vk mn} \left|g_{mn\nu}(\vk,\vq)\right|^2\delta(\epsilon_{n\vk}-\epsilon_{\rm F})\delta(\epsilon_{m\vk+\vq}-\epsilon_{\rm F}).
\end{multline}

\begin{figure*}[htp!]
    \centering
    \subfloat{\includegraphics{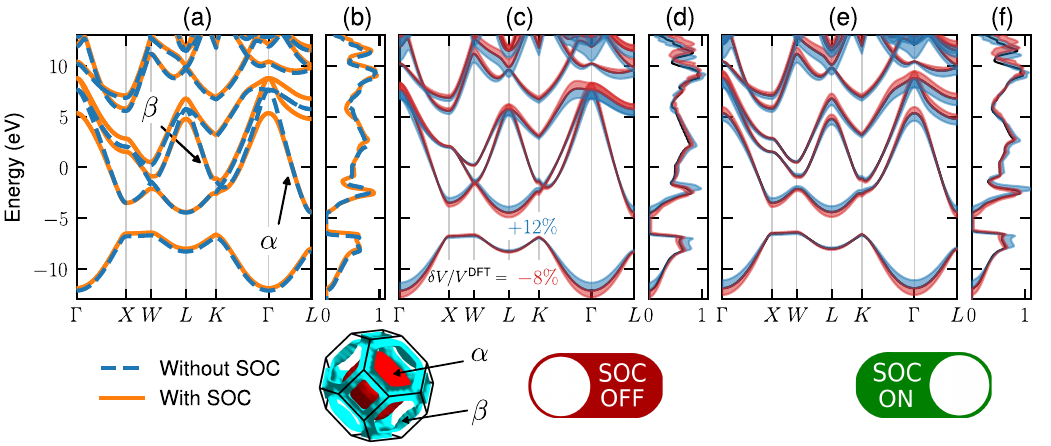}\label{subfig:band_structures_no_lattice_expansion}}
    \subfloat{\label{subfig:band_structures_dos_no_lattice_expansion}}
    \subfloat{\label{subfig:band_structures_lattice_expansion_nosoc}}
    \subfloat{\label{subfig:band_structures_dos_lattice_expansion_nosoc}}
    \subfloat{\label{subfig:band_structures_lattice_expansion_soc}}
    \subfloat{\label{subfig:band_structures_dos_lattice_expansion_soc}}
    \caption{\label{fig:band_structures}(a) Electronic band structures with (solid orange) and without (dashed blue) SOC computed at the DFT-optimized volume $\Vdft$. (b) The corresponding DOS. (c), (d) Band structure and DOS with varying unit cell volume going from $\delta V/\Vdft=\qtyrange[retain-explicit-plus]{-8}{+12}{\percent}$ without SOC. The shaded area represents where bands from intermediate volumes lie, and the solid curve in the center of the shaded area is for $\delta V/\Vdft=0$. The red shaded area depicts curves with decreasing volume, while the blue shaded area depicts curves of increasing volumes. (e), (f) Same as (c) and (d) with SOC. In all plots, the Fermi level is set at \num{0}. The $\alpha$ and $\beta$ Fermi surface sheets are shown at the bottom of panel (b) in red and blue respectively and are marked with an arrow in panel (a).}
\end{figure*}

\section{Computational Details}
\label{sec:computational_details}

The ground-state properties of Pb were calculated using DFT within a plane wave basis as implemented in the \textsc{Quantum Espresso} (QE) software package~\cite{giannozzi_quantum_2020}. 
We used an energy cut-off of \qty{90}{\rydberg}, a $(0.5~,0.5,~0.5)$ shifted \numproduct{36x36x36} $\vk$-point mesh and
a Gaussian smearing of \qty{1}{\milli\rydberg}.
To compute $F^{\rm el}$ we used a Fermi-Dirac distribution at the reported temperature.
We used DFT within the local density approximation (LDA)~\cite{perdew_self-interaction_1981} and used a fully relativistic ultrasoft~\cite{rappe_optimized_1990} pseudopotential from the PSLibrary version 0.2.2~\cite{dal_corso_pseudopotentials_2014}. 
The equilibrium volume at \qty{0}{\kelvin}, $\Vdft$, was obtained by optimizing the cell structure until the pressure was less than \qty{0.01}{\kilo\bar}. 
The fermi surface plot shown at the bottom of \cref{fig:band_structures} was generated using the \textsc{Fermi Surfer} program~\cite{kawamura_fermisurfer_2019}.
The phonon properties were computed using a coarse \numproduct{6x6x6} unshifted $\vq$-mesh from which the phonon dispersion and phonon free energies were obtained using Fourier interpolation. 
The phonon free energy was computed using frequencies interpolated on a finer \numproduct{64x64x64} $\vq$-mesh in Eq.~\eqref{eq:free_energy_phonons}.
Free energy calculations of \cref{eq:free_energy} sampled unit cell volume variations of \qtylist[retain-explicit-plus]{\pm 2;\pm 6;\pm 8;+10;+12}{\percent} around the optimized volume $\Vdft$.
The nesting function and phonon linewidth calculations of \cref{eq:nesting_function,eq:phonon_linewidth} were performed on a \numproduct{40x40x40} unshifted $\vk$-mesh and the delta functions were approximated by normalized Gaussian functions with a width of \qty{0.5}{\electronvolt}.
The calculations of transport properties (\cref{eq:ibte,eq:relaxation_time,eq:conductivity_tensor}) were carried out using the \textsc{EPW} code~\cite{ponce_epw:_2016,lee_electronphonon_2023} that enables the interpolation of all relevant quantities on fine grids via maximally localized Wannier functions~\cite{giustino_electron-phonon_2007,marzari_maximally_1997,souza_maximally_2001,pizzi_wannier90_2020}. 
Thus, the phonon frequencies $\omega_{\vq\nu}$, electronic eigenvalues $\epsilon_{n\vk}$, electron velocities $\vv_{n\vk}$ and EPC matrix elements $g_{mn\nu}(\vk,\vq)$ were computed on a \numproduct{60x60x60} $\vk$ and $\vq$ meshes. 
To speed up fine-grid interpolation, the coarse \numproduct{36x36x36} $\vk$ grid used to compute the SCF ground state was downsampled using a non-SCF calculation to an unshifted \numproduct{12x12x12} $\vk$-grid. 
Extensive convergence studies with respect to coarse and fine momentum grid sizes are given in the Supporting Information (SI)~\cite{Goudreault2024_SI}.
Finally, the computation of \cref{eq:ibte,eq:relaxation_time} requires the evaluation of Dirac delta distributions, which are approximated using the state-dependent Gaussian adaptive broadening approach detailed in Refs.~\cite{ponce_first-principles_2021_hall} and~\cite{li_shengbte_2014} which is implemented in \textsc{EPW} to avoid convergence studies related to these distributions.

\section{Results and discussion}
\label{sec:results}

\subsection{Ground state and volumetric effects}
\label{sec:equilibrium_properties}

\begin{figure*}[htp!]
    \centering
    \subfloat{\includegraphics{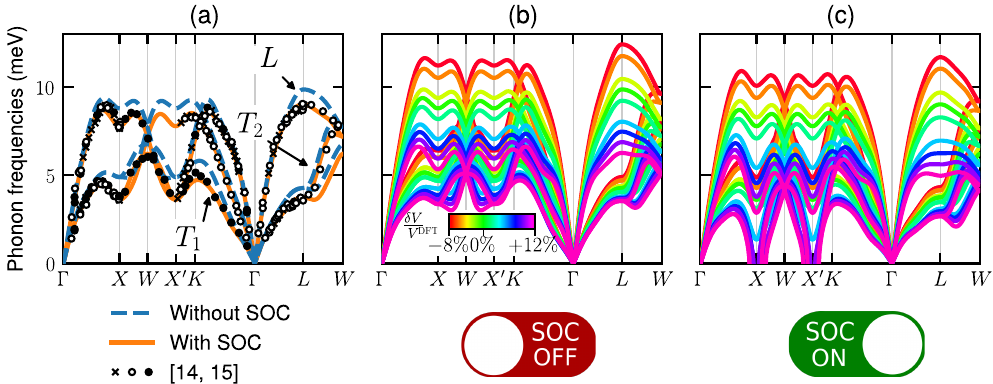}\label{subfig:phonon_dispersions_no_lattice_expansion}}
    \subfloat{\label{subfig:phonon_dispersions_lattice_expansion_nosoc}}
    \subfloat{\label{subfig:phonon_dispersions_lattice_expansion_soc}}
    \caption{\label{fig:phonon_dispersions}(a) Phonon dispersion of Pb with (solid orange) and without (dashed blue) SOC computed at the DFT-optimized lattice parameter. Markers represent neutron diffraction experimental data at \qty{100}{\kelvin} from Ref.~\cite{brockhouse_crystal_1962} ($\times$ markers) and from Ref.~\cite{brockhouse_old_1961} (open and closed circles). (b) Phonon dispersion of Pb without SOC at different volumes. The relative volume difference $\delta V/\Vdft\equiv (V-\Vdft)/\Vdft$ is denoted by the curve color where red corresponds to $\delta V/\Vdft =\qty{-8}{\percent}$ and magenta to $\delta V/\Vdft =\qty[retain-explicit-plus]{+12}{\percent}$. (c) Same as (b) with SOC.}
\end{figure*}

Geometry optimization yielded a lattice parameter of \qty{4.881}{\angstrom} without SOC and \qty{4.882}{\angstrom} with SOC. 
These values are within \qty{0.5}{\percent} of the experimental lattice parameter extrapolated at $T\rightarrow \qty{0}{\kelvin}$ which is \qty{4.905}{\angstrom}~\cite{grabowski_ab_2007}.
Consistent with previous reports~\cite{heid_effect_2010,corso_ab_2008,verstraete_density_2008}, SOC only changes the optimized lattice parameter by \qty{0.02}{\percent}. 
\Cref{subfig:band_structures_no_lattice_expansion,subfig:band_structures_dos_no_lattice_expansion} show the Fermi-level aligned band structure and DOS of Pb with and without SOC using their respective optimized lattice spacing. 
Pb having two bands crossing the Fermi level, the two resulting sheets in the Fermi Surface are denoted $\alpha$ and $\beta$, and are marked by arrows. Globally, SOC has little effect on the band structure and the DOS. 
The main differences can be seen in the band structure where the degeneracies are lifted close to the $\Gamma$ and $W$ points, which is consistent with what was previously reported in Refs.~\cite{heid_effect_2010,corso_ab_2008,verstraete_density_2008,horn_electronic_1984}. 
Notably, almost no changes occur around the Fermi level, which means that the Fermi surface is negligibly affected by SOC. However, we note that the slope of the bands near the Fermi level is slightly affected by SOC and these effects, albeit small as discussed in \cref{sec:transport}, have an influence on transport calculations.

\Cref{subfig:phonon_dispersions_no_lattice_expansion} shows the phonon dispersion of Pb with and without SOC using their respective optimized lattice parameter. 
The two transverse modes $T_1$ and $T_2$, as well as the longitudinal mode $L$ are identified by arrows. 
As reported by others~\cite{ponce_epw:_2016,corso_ab_2008,heid_effect_2010,verstraete_density_2008}, taking into account SOC softens the phonon frequencies, such that agreement with the inelastic neutron scattering data from Ref.~\cite{brockhouse_crystal_1962} is improved. 
Globally, this softening occurs almost everywhere in the Brillouin Zone and appears to be strongest near the $X$ point. 
Furthermore, SOC improves the agreement with the kinks in the dispersion attributed to Kohn anomalies~\cite{kohn_image_1959} on the $\Gamma-K$ line and close to the $L$ point.

For each volume variation listed in \cref{sec:computational_details}, we computed the ground state and phonon properties. 
The changes in the band structure and DOS of Pb using these variations are depicted in \cref{subfig:band_structures_lattice_expansion_nosoc,subfig:band_structures_dos_lattice_expansion_nosoc} without SOC and \cref{subfig:band_structures_lattice_expansion_soc,subfig:band_structures_dos_lattice_expansion_soc} with SOC. 
The shaded areas represent the change of eigenvalues upon a change in volume with respect to the Fermi energy. 
We observe that volume variations do not drastically change the band structures and have little to no effect on the DOS at the Fermi level. In general, a higher volume (blue) decreases the bandwidths and vice versa for decreasing volume (red). 
From this we conclude that both SOC and lattice variations have a negligible effect on the electronic structure close to the Fermi level.

The corresponding phonon dispersions are shown in \cref{subfig:phonon_dispersions_lattice_expansion_nosoc} without SOC and in \cref{subfig:phonon_dispersions_lattice_expansion_soc} with SOC. 
We see that the cell volume inversely correlates with the phonon frequencies, where smaller volumes yield higher frequencies, and vice versa. 
This is expected because a smaller volume generally gives higher interatomic force constants, which, in turn, increases the phonon frequencies. 
Hence, contrasting with the electronic band structure which is more sensitive to SOC than to the lattice parameter, the phonon dispersion is more sensitive to the volume than to SOC. 
The effect is so strong that the phonon softening induced by SOC at $X$ is enhanced to the point where the structure becomes unstable above an \qty{8}{\percent} dilation of the unit cell volume.  
This is seen in \cref{subfig:phonon_dispersions_lattice_expansion_soc} where the frequencies become imaginary in the vicinity of $\vq=X$. 
Still, Pb is known to be stable up to a temperature of \qty{600}{\kelvin} where its experimental variation in the volume of the unit cell is close to \qty{8}{\percent} with respect to its low temperature value~\cite{grabowski_ab_2007}. 
Therefore, such imaginary frequencies at these volumes should not occur, and we infer that there are possible mechanisms beyond the QHA that might stabilize the system like anharmonic lattice dynamics.
Indeed, anharmonic phonons were shown to have a strong effect at high temperatures. For instance, they are required in \ce{PbTe} as the usual harmonic approximation employed in DFPT fails at high temperature~\cite{romero_thermal_2015}. Furthermore, a recent work showed that anharmonic effects combined with a special displacement method were able to fix imaginary phonon modes obtained by DFPT in various other materials such as \ce{SrTiO3}~\cite{zacharias_anharmonic_2023}. Therefore, this avenue should be explored in future work in order to alleviate the phonon softening seen in \ce{Pb} with SOC near $\vq=X$.

\subsection{Fermi Surface Nesting and Phonon Linewidths}
\label{sec:nesting}

\begin{figure}%
    \centering
    \subfloat{\includegraphics{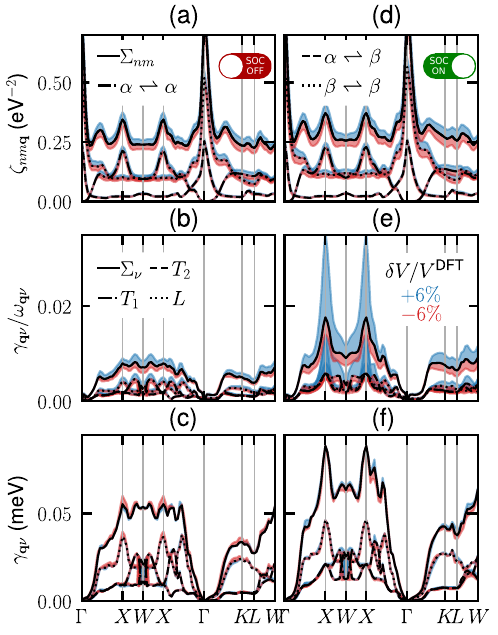}\label{subfig:nestings_zeta_nosoc}}
    \subfloat{\label{subfig:nestings_gzeta_nosoc}}
    \subfloat{\label{subfig:nestings_linewidth_nosoc}}
    \subfloat{\label{subfig:nestings_zeta_soc}}
    \subfloat{\label{subfig:nestings_gzeta_soc}}
    \subfloat{\label{subfig:nestings_linewidth_soc}}
    \caption{\label{fig:nestings}(a), (d) Electronic nesting function $\zeta_{\vq}$ from \cref{eq:nesting_function} where the dashed-dotted curve denotes the contribution from intraband transitions within the $\alpha$ sheet, the dotted curve within the $\beta$ sheet and the dashed curves denote the interband transitions between the two. The solid curve is the sum of all the band-resolved nestings. (c), (f) Phonon linewidths $\gamma_{\vq\nu}$ computed from \cref{eq:phonon_linewidth} where the dashed-dotted and dashed curves represent contributions from the transverse modes $T_1$ and $T_2$ and the dotted curve represent the longitudinal mode $L$. The solid curves represent the sum of all the mode-resolved linewidths. (b), (e) $g$-weighted nesting function $\gamma_{\vq\nu}/\omega_{\vq\nu}$. Mode resolved contributions are depicted with the same line styles as for (c). (a-f) All black lines are computed at the DFT-optimized volume $\Vdft$ while the blue shaded area spans the curves with increasing volumes up to $\delta V/\Vdft=\qty[retain-explicit-plus]{+6}{\percent}$ while the red shaded area represents decreasing volumes $\delta V/\Vdft=\qty{-6}{\percent}$. (d), (e) and (f) include SOC while (a), (b) and (c) do not. 
    }
\end{figure}

To shed light on the SOC-induced phonon softening in Pb, we computed the Fermi surface nesting function using \cref{eq:nesting_function} with its band-resolved version $\zeta_{nm\vq}$. 
A detailed study of $\zeta_{nm\vq}$ without SOC was already reported by \citet{rittweger_phonon_2017} where it was shown that both interband and intraband transitions play an important role in the nesting. 
In this section, we expand this analysis with SOC, the EPC matrix elements and volume variations.

The nesting function is shown in \cref{subfig:nestings_zeta_nosoc,subfig:nestings_zeta_soc} respectively without SOC and with SOC. 
As for \cref{fig:band_structures,fig:phonon_dispersions}, the shaded areas represent the regions spanned by the curves computed by expanding (blue) or compressing (red) the volume. 
We distinguish between the intraband nesting for the $\alpha$ sheet (dashed-dotted curve) and the $\beta$ sheet (dotted curve), the interband nesting between the two sheets (dashed curve) and the total sum of all of these (plain curve). 
We observe that the total nesting function increases with SOC by a factor of \qtyrange{5}{15}{\percent} depending on $\vq$. 
Furthermore, the nesting does indeed peak at $\vq =X$, which explains why the phonon softening at this point is higher. 
From the band resolved curves, we see that this peak is mainly due to interband nesting, whereas the intraband nesting is mostly constant across the Brillouin Zone except in the vicinity of $\Gamma$ where it diverges in an integrable way. 
Also, the intraband transitions are much larger with the $\beta$ sheet than for the $\alpha$ sheet, which is even lower than the interband nesting. 
This is interesting, as the shape of the $\beta$ sheet is quite complicated and anisotropic with its tubular geometry near the edge of the Brillouin Zone~\cite{rittweger_phonon_2017} compared to the $\alpha$ sheet, which holds seemingly large parallel surface areas~\cite{corso_ab_2008,rittweger_phonon_2017}. 
As such, intuitively, one could have expected that the nesting within the $\alpha$ sheet would be higher than within the $\beta$ sheet. 
Regardless, changing the volume has quantitative effects on the nesting where an increase in volume increases the nesting and vice versa.

On the other hand, as mentioned in \cref{subsec:fs_nesting_ph_linewidth}, the nesting function only contains part of the information and one needs to take into account the EPC transition probabilities in order to analyze phonon softening.
\Cref{subfig:nestings_linewidth_nosoc,subfig:nestings_linewidth_soc} show the mode-resolved phonon linewidth from \cref{eq:phonon_linewidth} (dashed, dashed-dotted and dotted curves) and their sum (solid lines) respectively, without SOC and with it. 
We observe that the total linewidth with SOC is much higher than without it, which implies that the phonon lifetimes are lower with SOC. 
Also, we see that the phonon linewidth without SOC does not peak at $X$ compared to SOC, where it strongly peaks at this point. 
As such, we conclude that the EPC matrix element is enhanced at $X$ specifically with SOC. 
In particular, the peak at $X$ with SOC is due to the linewidth of the transverse modes $T_1$ (dashed-dotted curve) and $T_2$ (dashed curve) where a peak suddenly appears with SOC while the longitudinal mode $L$, which already peaks at $X$ without SOC, has its peak enhanced with SOC. 
Furthermore, there is a global trend upward throughout the Brillouin Zone of the linewidth for all modes, which correlates with the global phonon softening. 
As for the volumetric effect, it is almost non-existent for the linewidth which implies that the volume dependence of the EPC matrix elements is mostly contained within the phonon frequencies. 
Indeed, this is seen in \cref{subfig:nestings_gzeta_nosoc,subfig:nestings_gzeta_soc} where we show the phonon linewidth divided by their corresponding phonon frequencies, namely $\gamma_{\vq\nu}/\omega_{\vq\nu}$. 
This is equivalent to a nesting function weighted by the EPC matrix element, and doing so exposes the effect of the frequencies changes onto the matrix element as $g$ scales with $\omega_{\vq\nu}^{-1/2}$. 
Due to this relation and the fact that phonon frequencies strongly soften with SOC, the volume dependence of $\gamma_{\vq\nu}/\omega_{\vq\nu}$ increases with expanding volume with SOC (see the blue-shaded area in \cref{subfig:nestings_gzeta_soc}). 
Without SOC (\cref{subfig:nestings_gzeta_nosoc}), the volume dependence is much smaller and similar to the volume changes seen in the nesting function (\cref{subfig:nestings_zeta_nosoc}). 
As for the linewidth, the peak at $X$ remains with SOC while it is flattened without SOC.

In summary, electronic nesting does indeed peak at $X$, which is due to interband transitions (\cref{subfig:nestings_zeta_nosoc,subfig:nestings_zeta_soc}), and this is seen both with and without SOC. 
However, this nesting peak is flattened out by the EPC matrix element without SOC (\cref{subfig:nestings_gzeta_nosoc,subfig:nestings_linewidth_nosoc}), while it is enhanced with SOC (\cref{subfig:nestings_gzeta_soc,subfig:nestings_linewidth_soc}). 
This suggests that the softening of the phonon frequencies is stronger at $X$ because the EPC matrix element becomes larger.
Simultaneously, the enhanced phonon softening with increasing volume affects only the frequency part of the EPC matrix element, which makes it even stronger. 
As such, we conclude that there is a mutual interplay between the phonon softening and the EPC matrix element, which
 yields the observed dispersion. 
Therefore, from our observations we conclude that not only nesting is important to explain the phonon softening in Pb but also that the EPC matrix elements play a role in this phenomenon.

\subsection{Free Energy and thermal expansion}
\label{sec:free_energy}

\begin{figure}
    \centering
    \subfloat{\includegraphics{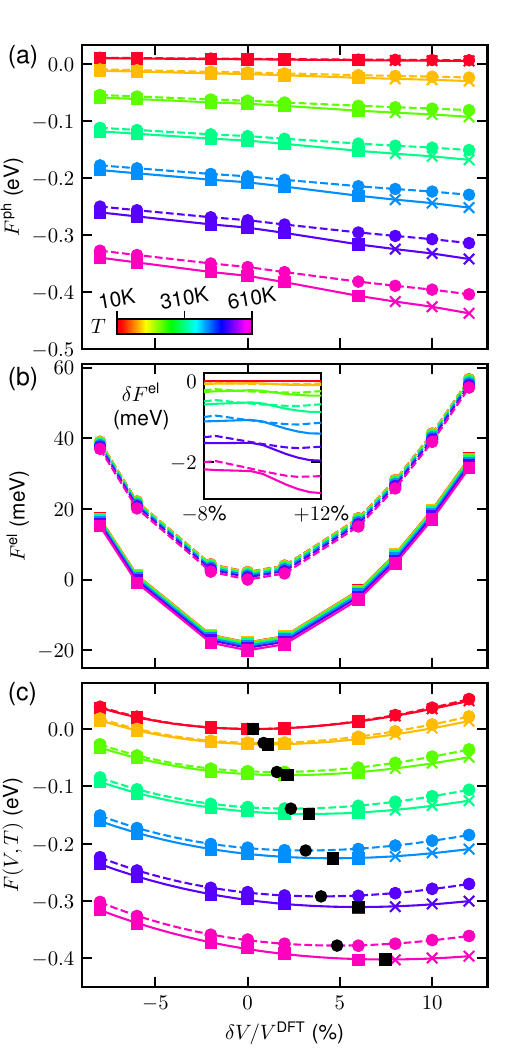}\label{subfig:free_energy_phonons}}
    \subfloat{\label{subfig:free_energy_electrons}}
    \subfloat{\label{subfig:free_energy_total}}
    \caption{\label{fig:free_energy}(a) $\Fph$ as a function of $\delta V/\Vdft$ and $T$. (b) $\Fel$ as a function of $\delta V/\Vdft$ and $T$. The curves were shifted vertically such that $F(\Vdft,T=\qty{610}{\kelvin})=0$ without SOC and \qty{-20}{\milli\electronvolt} with SOC. The inset shows $\delta\Fel=\Fel(V, T)-\Fel(V,T=\qty{10}{\kelvin})$. (c) Total free enerSERTAgy $F=\Fel+\Fph$ as a function of $\delta V/\Vdft$ and $T$. The dashed (without SOC) and solid (with SOC) lines are fits to \cref{eq:birch-murnaghan}. The free energy is shifted vertically so that $F(\Vdft, T=\qty{10}{\kelvin})=0$. The black rounded (resp. squared) markers signals the position of $V_0(T)$ as extracted from the fit without SOC (resp. with SOC). (a)-(c) Dashed curves with square or $\times$ markers include SOC, while the plain ones with round markers do not. The $\times$ markers denote points where negative phonon frequencies occurred.
    }
\end{figure}
From the volume-dependent phonon frequencies calculated, we used \cref{eq:free_energy_phonons} to obtain the phonon contribution to the free energy at various temperatures.
The results are shown in \cref{subfig:free_energy_phonons} where the solid curves include SOC while the dashed ones do not.
The $\times$ markers denote values where some phonon frequencies become imaginary.
Such frequencies were simply discarded when using \cref{eq:free_energy_phonons}.
As expected, $\Fph$ decreases both if the temperature or volume increases.

The electronic contribution to the free energy is shown in \cref{subfig:free_energy_electrons} for different temperatures and volumes.
Again, the solid curves include SOC while the dashed ones do not. 
It is clear from this plot that most of the variation in the electronic free energy comes from changes in the lattice parameter.
This is shown by subtracting the free energy at the lowest temperature for each curve: $\delta \Fel = \Fel(V, T) - \Fel(V, T=\qty{10}{\kelvin})$, as depicted in the inset.
Doing so effectively cancels out the variation due to the lattice parameter, and the remainder only shows a temperature variation of less than \qty{3}{\milli\electronvolt}, which is roughly two orders of magnitude smaller than temperature variations in $\Fph$.
As such, we conclude that, although \ce{Pb} is metallic, the term $\Felex$ is negligible compared to $\Fph$.

Finally, \cref{subfig:free_energy_total} shows the total free energy for different temperatures with respect to the sampled volumes.
The solid and dashed curves are fits to the third-order Birch-Murnaghan EOS (\cref{eq:birch-murnaghan}) respectively with and without SOC.
As for \cref{subfig:free_energy_phonons}, $\times$ markers denote points where negative phonon frequencies occurred. The black dots and squares denote, respectively, the $(V_0, F_0)$ pairs extracted from the fit for each temperature with and without SOC. 

\begin{figure}
    \centering
    \subfloat{\includegraphics{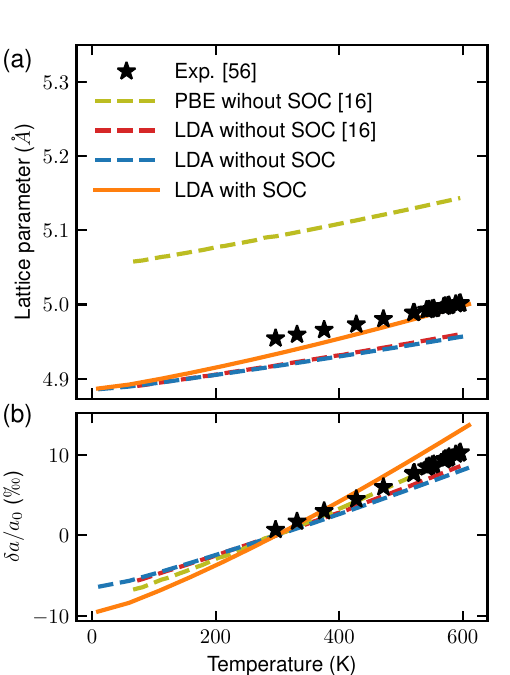}\label{subfig:lattice_expansion_absolute}}
    \subfloat{\label{subfig:lattice_expansion_relative}}
    \caption{\label{fig:lattice_expansion}
    (a) Lattice parameter as a function of temperature extracted from the free energy minimum $V_0$. (b) Lattice parameter relative variation $\delta a/a_0$ where $a_0$ is the lattice parameter at a reference temperature of $T=\qty{298}{\kelvin}$. Experimental data from Ref.~\cite{feder_use_1958} is shown as black stars. (a)-(b) The solid orange curve includes SOC, while the dashed blue one does not. DFT results without SOC using LDA (dashed red) and PBE (dashed yellow) from Ref.~\cite{grabowski_ab_2007} are shown for comparison.
    }
\end{figure}

\Cref{subfig:lattice_expansion_absolute} reports the values of the lattice parameter obtained from the EOS fits as a function of temperature. 
For comparison, we show the DFT results without SOC obtained in Ref.~\cite{grabowski_ab_2007} with a red dashed line. 
Our results closely match theirs with LDA.
Further comparison is also possible with experimental measurements from Ref.~\cite{feder_use_1958} denoted by black stars. 
We observe that SOC induces a higher absolute variation in the lattice parameter than without SOC, which places the equilibrium lattice parameter in better agreement with the experimental measurements at high temperature.
Nonetheless, equilibrium volumes both with and without SOC remain lower than the experimental data, which is consistent with the fact that LDA typically underestimates the lattice parameter. 
PBE results from Ref.~\cite{grabowski_ab_2007} are indicated with a yellow line; they overestimate the equilibrium lattice parameter for all temperatures as it is also expected for this functional. 
The relative temperature variation $\delta a/a_0=[a(T)-a_0]/a_0$ is shown in \cref{subfig:lattice_expansion_relative} where $a_0=a(T=\qty{298}{\kelvin})$.
Without SOC, we observe that the slope is smaller than the experimental variation, while with SOC it is larger. 
In comparison, the PBE variation (without SOC) is the best matched to experimental measurements for this variation. 
To our knowledge, there are no reports on the temperature variation of the equilibrium lattice parameter computed using PBE with SOC for Pb.

\subsection{Heat Capacity and bulk modulus}
\label{subsec:results_heat_capacity}

\begin{figure}
    \centering
    \includegraphics{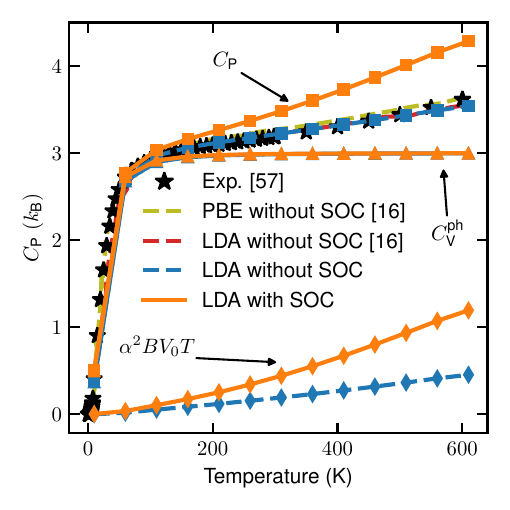}
    \caption{\label{fig:heat_capacity}Decomposition of the heat capacity of \ce{Pb} at constant pressure in function of temperature. The solid orange curves include SOC while the dashed blue ones do not. Square markers denote SERTAthe total heat capacity at constant pressure as computed from \cref{eq:heat_capacity_constant_pressure}, triangles represent the contribution from $\CVph$ and diamonds show the contribution from the $\alpha^2BV_0T$ term. We compare with experimental data from Ref.~\cite{arblaster_thermodynamic_2012} shown as black stars and DFT results without SOC from Ref.~\cite{grabowski_ab_2007} computed using the LDA (red dashed curve) and the PBE (yellow dashed curve) exchange-correlation functionals.}
\end{figure}

\Cref{fig:heat_capacity} shows the heat capacity as computed from \cref{eq:heat_capacity_constant_pressure} using the equilibrium lattice volume $V_0$ and the isothermal bulk modulus $B$ extracted from the EOS fits.
The dashed blue curves denote our results without SOC, while the solid orange curves include it.
The total heat capacity at constant pressure as computed from \cref{eq:heat_capacity_constant_pressure} is shown using square markers.
We also show the different contributions to \cref{eq:heat_capacity_constant_pressure} where triangles represent the $\CVph$ term and diamonds denote the $\alpha^2 BV_0(T)T$ offset.
We also compare with previous reported DFT results computed without SOC by \citet{grabowski_ab_2007} and experimental data from Ref.~\cite{arblaster_thermodynamic_2012}.
Our results without SOC are in excellent agreement with experimental and reference DFT data, while results with SOC overestimate the experimental data at high temperature.
This is surprising, as both the phonon dispersion and TE are in better agreement with SOC than without SOC.
This difference is mainly imputable to the last term of the right-hand side of \cref{eq:heat_capacity_constant_pressure} where $\alpha$, $V_0$ and $B$ are all higher with SOC than without it, while $\CVph$ is almost the same with and without SOC. Meanwhile, the $\CVel$ term (not shown) is less than \qty{e-2}~$\kB${} at \qty{600}{\kelvin} both with and without SOC and is thus negligible.
It is interesting to note that the reported PBE results without SOC display better agreement with experimental data than our LDA results with SOC. 
%
As reported results so far showed good agreement between energy functionals, a better description of the electronic structure might be needed in order to improve results with SOC. For instance, the GW formalism might be worth investigating as it often provides better band structures, ground state and phonon related properties with respect to experiments compared to bare DFT calculations~\cite{ponce_towards_2018,antonius_many-body_2014,li_electron-phonon_2019,kutepov_ground-state_2009,filip_gw_2014}. 
This suggests that GW could yield better heat capacity results when coupled with SOC. However, such investigations are beyond the scope of this work. 

\begin{figure}
    \subfloat{\includegraphics{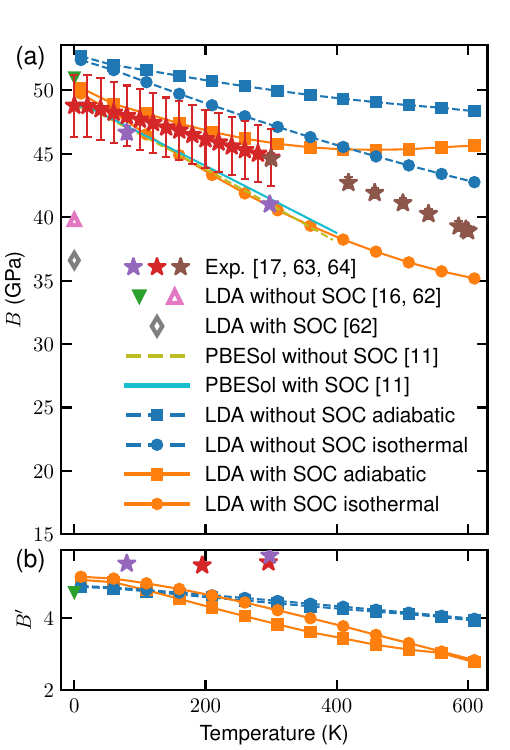}\label{subfig:bulk_modulus}}
    \subfloat{\label{subfig:bulk_modulus_derivative}}
    \caption{\label{fig:bulk_modulus}(a) Bulk modulus of \ce{Pb}. (b) The pressure derivative of the bulk modulus. (a)-(b) The dashed blue and solid orange lines are results from this work respectively without and with SOC. Round markers represent the isothermal bulk modulus $B$ (a) or its pressure derivative $B'$ (b) as extracted from the third order Birch-Murnaghan fit of the free energy from the left panel of \cref{fig:lattice_expansion}. Meanwhile, square markers denote the adiabatic bulk modulus $\BA$ (a) or its pressure derivative $\BAprime$ (b) computed from \cref{eq:theory_adiabatic_bulk_modulus,eq:theory_adiabatic_bulk_modulus_derivative,eq:theory_adiabatic_bulk_modulus_derivative_offset}. The dashed yellow and teal lines represent the isothermal bulk modulus computed using the PBESol functional from Ref.~\cite{smirnov_effect_2018}. Other LDA results computed at $T=\qty{0}{\kelvin}$ are from Refs.~\cite{grabowski_ab_2007} (green triangle) and~\cite{dal_corso_projector_2012} (pink open triangle without SOC and gray open diamond with SOC). Isothermal experimental data is from Ref.~\cite{strassle_equation_2014} (purple stars) while adiabatic experimental data is from Refs.~\cite{miller_pressure_1969} (red stars) and \cite{vold_elastic_1977} (brown stars).
    }
\end{figure}

\Cref{subfig:bulk_modulus,subfig:bulk_modulus_derivative} show the temperature dependence of the adiabatic and isothermal bulk modulus and their pressure derivative calculated using \cref{eq:theory_adiabatic_bulk_modulus,eq:theory_adiabatic_bulk_modulus_derivative,eq:theory_adiabatic_bulk_modulus_derivative_offset}.
The orange solid curves include SOC, while the dashed blue ones do not.
Curves with square markers represent the adiabatic bulk modulus, whereas the ones with round markers represent the isothermal bulk modulus and the same goes for their pressure derivative.
We compare our results with experimental data taken from Refs.~\cite{waldorf_low-temperature_1962,vold_elastic_1977} for $\BA$ (red and brown stars), Ref.~\cite{strassle_equation_2014} for $B$ and $B'$ (purple stars) and from Ref.~\cite{miller_pressure_1969} for $\BAprime$. We also compare with DFT calculations from Refs.~\cite{grabowski_ab_2007,dal_corso_projector_2012} (green and pink triangles and gray diamond) and from isothermal calculations by Ref.~\cite{smirnov_effect_2018} (yellow and teal lines).
To our knowledge, there are no reports of DFT calculations of the full temperature dependence of the adiabatic bulk modulus and its pressure derivative for \ce{Pb} both with and without SOC.
Our results with SOC are in good agreement with experimental data at low temperature both for $B$ and $\BA$ whereas, our results show that neglecting SOC yields an overestimated adiabatic and isothermal bulk modulus. Furthermore, our isothermal results with SOC are in good agreement with previous PBEsol results from \citet{smirnov_effect_2018}.
The high-temperature discrepancy with SOC is probably due to the one seen in the heat capacity as $\BA$ is directly proportional to it, see \cref{eq:theory_adiabatic_bulk_modulus}.
Compared to other $T=\qty{0}{\kelvin}$ LDA calculations, both of our low-temperature results for $\BA$ and $\BAprime$ agree well with the ones from~\cite{grabowski_ab_2007} while the difference is much larger for Ref.~\cite{dal_corso_projector_2012}, although the latter did not use a free energy description.
In the end, as for the phonon dispersion, SOC seems to soften both the isothermal and adiabatic bulk modulus.

As for the bulk modulus pressure derivative depicted in \cref{subfig:bulk_modulus_derivative}, we observe a strong deviation from the experimental measurements where our results for $B'$ and $\BAprime$ decrease with temperature while the experiments show an increase. Both $B'$ and $\BAprime$ are also relatively close to each other at all temperatures with and without SOC which qualitatively agrees with experimental results and indicates that pressure has a similar effect on both of $B$ and $\BA$.
Although it has a different temperature trend from the experimental measurements, the overall quantity is positive for all temperatures, which means that an increase in pressure would lead to an increase of the bulk modulus.
Our low temperature results are also in good agreement with the previous $T=\qty{0}{\kelvin}$ result from Ref.~\cite{grabowski_ab_2007} without SOC.
We see that including SOC causes $B'$ to decrease faster with temperature than without it.
As for heat capacity calculations, GW quasiparticle corrections could be employed in order to have a more accurate description of the electronic structure and free energies. These effects could change the temperature trends of the bulk modulus and its pressure derivative both with and without SOC. In addition, taking into account anharmonic phonon contributions might affect the TE, the heat capacity and the bulk modulus results. Such contributions were computed in \ce{Pb} without SOC and a non negligible positive contribution to the free energy and a negative one to the heat capacity were found~\cite{glensk_understanding_2015}. If these corrections carries out to a calculation with SOC, the agreement with experimental data might improve. 

\subsection{Charged carrier transport}
\label{sec:transport}

\begin{figure}
    \centering
    \subfloat{\includegraphics{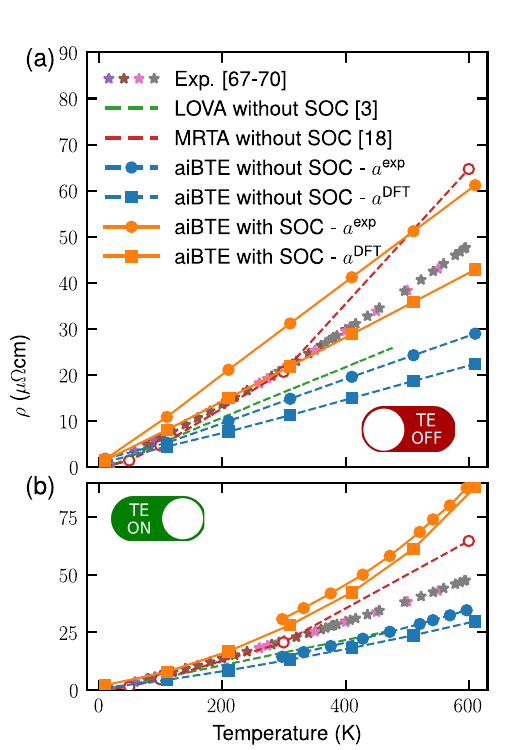}\label{subfig:transport_no_te}}
    \subfloat{\label{subfig:transport_with_te}}
    \caption{\label{fig:transport_results}(a) aiBTE results without thermal expansion. The solid orange curves include SOC while the dashed blue ones do not. Curves with square markers were computed using the optimized geometry $\Vdft$ while the ones with round markers used the experimental lattice parameter at room temperature of $a^{\text{exp}}=\qty{4.95}{\angstrom}$ taken from Ref.~\cite{zagorac_recent_2019}. The purple, pink, gray and brown stars represent the experimentally measured resistivity from Refs.~\cite{hellwege_electrical_1982,moore_absolute_1973,cook_thermal_1974,leadbetter_energy_1966}. The green dashed curve is the resistivity obtained using the LOVA approach as reported in Ref.~\cite{savrasov_electron-phonon_1996} while the red dashed line represents MRTA results obtained with the experimental lattice parameter as reported in Ref.~\cite{rittweger_phonon_2017}. Both the green and red dashed curve neglect SOC. (b) Same as (a) except TE is included in the computation of the solid orange and dashed blue curves. Curves with squared markers make use of the TE computed from the minimum of the free energy $V_0(T)$ while the ones with round markers used the experimental lattice parameters taken from Ref.~\cite{feder_use_1958}.
    }
\end{figure}
In this section, we report our results for the phonon-limited temperature-dependent resistivity of Pb as computed using \cref{eq:conductivity_tensor,eq:ibte,eq:relaxation_time}.
The results are shown in \cref{subfig:transport_no_te} without TE and in \cref{subfig:transport_with_te} with it.
When TE is included, we compute a resistivity of $\rho=\qty{13.2}{\micro\ohm\centi\meter}$ without SOC and $\rho=\qty{28.2}{\micro\ohm\centi\meter}$ with SOC at $T=\qty{310}{\kelvin}$.
When neglecting TE, we obtain $\rho=\qty{11.3}{\micro\ohm\centi\meter}$ without SOC and $\rho=\qty{21.9}{\micro\ohm\centi\meter}$ with it.
The experimental value at this temperature is $\rho=\qty{22.2}{\micro\ohm\centi\meter}$~\cite{moore_absolute_1973}.
For comparison, we also show earlier results by \citet{savrasov_electron-phonon_1996} (green dashed curve) and \citet{rittweger_phonon_2017} (red dashed curve with rounded markers), where the former used Allen's formula (LOVA)~\cite{allen_new_1978} and the latter used MRTA~\cite{ponce_first-principles_2020,claes_assessing_2022} without SOC.
In MRTA, the second term of \cref{eq:ibte} is neglected and a geometric factor of $\left(1-\frac{\vv_{n\vk}\cdot\vv_{m\vk+\vq}}{|\vv_{n\vk}|^2}\right)$ is inserted within the integral of \cref{eq:relaxation_time}.
For both the red and green dashed lines, SOC was neglected.
We compare against the experimental data from Refs.~\cite{hellwege_electrical_1982,leadbetter_energy_1966,cook_thermal_1974,moore_absolute_1973} which are represented by the purple, pink, gray and brown stars.
For both panels, solid orange curves include SOC, whereas the dashed blue ones do not.
Curves with square markers were computed using DFT geometry obtained from relaxation for \cref{subfig:transport_no_te} or from minimization of the free energy at each temperature, as shown in \cref{subfig:transport_with_te}.
In contrast, the curves with round markers were computed using the experimental lattice parameter at room temperature taken from Ref.~\cite{zagorac_recent_2019} in \cref{subfig:transport_no_te} and the experimental temperature-dependent lattice parameters of Ref.~\cite{feder_use_1958} in \cref{subfig:transport_with_te}.
We observe that curves computed using the experimental lattice parameter (with round markers) all have higher resistivity than their corresponding DFT analogs.
This is expected as the experimental volume is higher for all temperatures, yielding softer phonons and increased EPC matrix elements.
As such, this increases the probability that an electron diffuses via a phonon, thus decreasing the mean free path of electrons, which increases the resistivity.
The same can be said when SOC is turned on (solid orange curves), where, in all cases, the resistivity increases.
This was also seen using the LOVA approach, as reported in Refs.~\cite{ponce_epw:_2016,smirnov_effect_2018} and is consistent with the strong phonon softening observed with SOC, which increases the EPC matrix element, as discussed in \cref{subsec:fs_nesting_ph_linewidth}.

Focusing on \cref{subfig:transport_no_te}, the resistivity results obtained are all linear with the temperature at high temperature.
This is expected as the occupation number of phonons $n_{\vq\nu}\propto T$ when $\kB T\gg \hbar\omega_{\vq\nu}$ in \cref{eq:ibte,eq:relaxation_time}.
We see that, as for the LOVA approach~\cite{ponce_epw:_2016,smirnov_effect_2018}, SOC needs to be taken into account to have values close to the experimental data.
Nevertheless, this is still insufficient to explain the small deviation from linearity at high temperature, and thus we conclude that the aiBTE is unable to explain the discrepancy seen in previous reports.
When introducing TE, we see that DFT results without SOC do indeed get a small non-linear component at high temperature but, with SOC, this effect is much stronger which makes the resistivity increase rapidly at high temperature.

In addition, the aiBTE formalism gives a slightly different result than the so-called Self-Energy Relaxation Time Approximation (SERTA) approach where the second term on the right hand side of \cref{eq:ibte} is neglected.
At $T=\qty{310}{\kelvin}$, the SERTA overestimates the resistivity with TE by \qty{14.7}{\percent} without SOC and by \qty{5.7}{\percent} with SOC compared to the aiBTE. Without TE, the relative difference is similar where the SERTA overestimates the resistivity by \qty{16.1}{\percent} without SOC and \qty{7.0}{\percent} with SOC compared to the aiBTE.
See the Supporting Information for a comparison of the SERTA and aiBTE with temperature~\cite{Goudreault2024_SI}.
\begin{figure}
    \centering
    \subfloat{\includegraphics{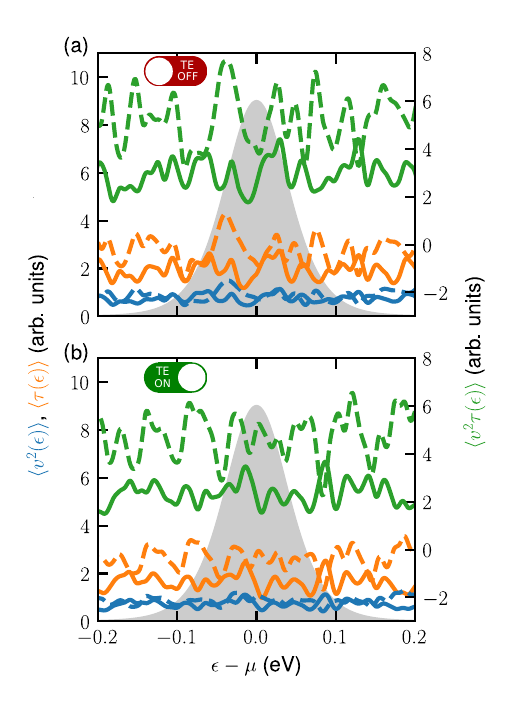}\label{subfig:v2_tau_v2tau_fs_average_no_te}}
    \subfloat{\label{subfig:v2_tau_v2tau_fs_average_with_te}}
    \caption{Fermi surface averages of the squared velocity matrix elements $\left\langle v^2(\epsilon)\right\rangle$ in blue, the relaxation times $\left\langle\tau(\epsilon)\right\rangle$ in orange and their product $\left\langle v^2\tau(\epsilon)\right\rangle$ in green. See text for the corresponding equations. Solid lines include SOC while dashed lines do not. All curves are computed at $T=\qty{310}{\kelvin}$. The gray shaded area represents $-\frac{df^0}{d\epsilon}$ where $f^0$ is the Fermi-Dirac occupation function at the same temperature. (a) Without TE. (b) With TE. All delta functions are approximated as Gaussians with a width of \qty{5}{\milli\electronvolt}.
    \label{fig:v2_tau_v2tau_fs_average}}
\end{figure}

It is worth noting that our resistivity results with SOC using the experimental lattice parameter is higher than experimental measurements which is quite unusual (solid orange curve with round markers in \cref{subfig:transport_no_te}). In fact, typical resistivity calculations employing the aiBTE formalism usually underestimate the experimental data. Since the results with $a^{\text{DFT}}$ and SOC are quite close to experiments and because the EPC matrix elements with SOC are quite sensitive to geometry variations as depicted in \cref{subfig:nestings_gzeta_soc}, we argue that this substantive increase of the lattice parameter might cause such a gain which is reflected into the resistivity through the relaxation times as per \cref{eq:conductivity_tensor,eq:ibte,eq:relaxation_time}.
This is supported by a similar trend when TE is taken into account in \cref{subfig:transport_with_te} where the temperature-dependent lattice parameters with SOC are much higher than the optimized one at high temperature (see \cref{fig:lattice_expansion}). However, the velocity matrix elements also need to be taken into account into the analysis since they are needed to compute the conductivity in \cref{eq:conductivity_tensor}. Despite the fact that the bands barely change with SOC, their slope does change slightly.
The difference can be measured via a Fermi surface average of the squared velocity matrix elements $\left\langle v^2(\epsilon)\right\rangle=\frac{1}{N_k}\sum_{n\vk}\left\|\vv_{n\vk}\right\|^2\delta(\epsilon-\epsilon_{n\vk})$ which is depicted by the blue curves in \cref{subfig:v2_tau_v2tau_fs_average_no_te} without TE and \cref{subfig:v2_tau_v2tau_fs_average_with_te} with TE at $T=\qty{310}{\kelvin}$. In these figures, solid curves include SOC while dashed ones do not. We observe that SOC does indeed decrease $\left\langle v^2(\epsilon)\right\rangle$ at the Fermi level by $\sim\qty{15}{\percent}$ without TE and $\sim\qty{35}{\percent}$ with TE at this temperature.
For comparison, we compute the relaxation times averaged on the Fermi surface $\left\langle \tau(\epsilon)\right\rangle=\frac{1}{N_k}\sum_{n\vk}\tau_{n\vk}\delta(\epsilon-\epsilon_{n\vk})$ which is depicted by the orange curves in \cref{fig:v2_tau_v2tau_fs_average}. We see that SOC induces a decrease of $\sim\qty{20}{\percent}$ without TE and $\sim\qty{50}{\percent}$ with it at the Fermi level for the same temperature. In order to see the combined effect of these decreases, we compute the averaged product of these functions on the Fermi surface using $\left\langle v^2\tau(\epsilon)\right\rangle = \frac{1}{N_k}\sum_{n\vk}\left\|\vv_{n\vk}\right\|^2\tau_{n\vk}\delta(\epsilon-\epsilon_{n\vk})$ which amounts to compute the first term of \cref{eq:ibte} without the derivative of the occupation factor $-\frac{df^0}{d\epsilon}$, the latter being depicted by the gray shaded area in \cref{fig:v2_tau_v2tau_fs_average}. Doing so brings forward the total effect of SOC where $\left\langle v^2\tau(\epsilon)\right\rangle$ is decreased by $\sim\qty{38}{\percent}$ without TE and $\sim\qty{56}{\percent}$ with it at the Fermi level at the same temperature. We thus see that the SOC effect on the squared velocities and relaxation times adds up to yield a higher difference when these functions are combined both with TE and without it. We notice that the total decrease is closer to the decrease of the relaxation times than the squared velocities which suggests that the final resistivity calculation is more sensitive to the former than the latter.

\begin{figure}
    \centering
    \subfloat{
    \includegraphics{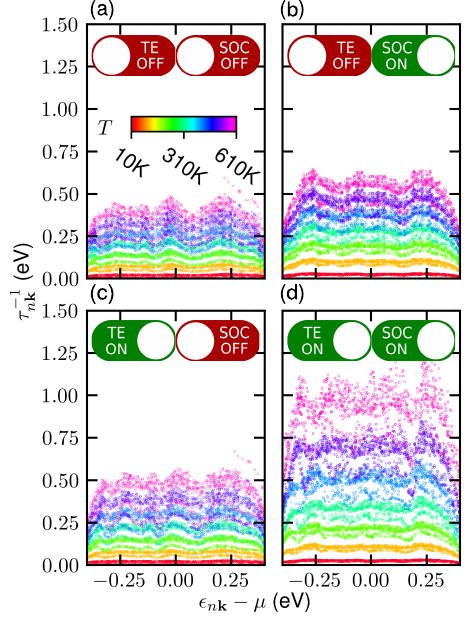}\label{subfig:relaxation_times_no_te_nosoc}}
    \subfloat{\label{subfig:relaxation_times_no_te_soc}}
    \subfloat{\label{subfig:relaxation_times_te_nosoc}}
    \subfloat{\label{subfig:relaxation_times_te_soc}}
    \caption{\label{fig:relaxation_times}EPC relaxation times inverse  as computed from \cref{eq:relaxation_time} in function of the difference between the corresponding single-particle energy eigenvalues $\epsilon_{n\vk}$ and the chemical potential $\mu$. (b) and (d) include SOC while (a) and  (c) do not. (c) and  (d) include TE while (a) and (b) do not. It is worth nothing here that $\mu$ is temperature-dependant.}
\end{figure}

To shed light on the anomalously high resistivity with SOC and TE, in \cref{fig:relaxation_times} we show the inverse relaxation times resolved by temperature computed from \cref{eq:relaxation_time} as a function of their corresponding electronic eigenvalue $\epsilon_{n\vk}$.
The results without SOC are shown in \cref{subfig:relaxation_times_no_te_nosoc,subfig:relaxation_times_te_nosoc}, while those with SOC are shown in \cref{subfig:relaxation_times_no_te_soc,subfig:relaxation_times_te_soc}. 
Meanwhile, the results without TE are depicted in \cref{subfig:relaxation_times_no_te_nosoc,subfig:relaxation_times_no_te_soc} and those with TE are shown in \cref{subfig:relaxation_times_te_nosoc,subfig:relaxation_times_te_soc}. 
In all cases, increasing the temperature decreases the relaxation times, which in turn increases the resistivity. This is due to the increased phonon population, as described in a previous paragraph. 
When SOC is turned on, we see that the inverse relaxation time increases by about \qty{\sim 25}{\percent} at high temperature before TE is considered. 
The effect of TE without SOC has a similar effect to that of SOC without TE and increases the inverse relaxation times by \qty{\sim 10}{\percent} at high temperature. 
However, the effect of TE with SOC is more drastic as it yields an increase of \qty{\sim 75}{\percent}, thus tripling the effect of SOC alone.
This signals an interplay between SOC and TE with respect to relaxation times, and these behaviors are correlated with those seen in $\gamma_{\vq\nu}/\omega_{\vq\nu}$ depicted in \cref{subfig:nestings_gzeta_soc}. 
Indeed, the phonon linewidth is increased by SOC but is independent of the volume, while the phonon frequencies are sensitive to it. 
As such, EPC matrix elements experience a strong increase with volume because the phonon dispersion is significantly softened by it (see \cref{subfig:phonon_dispersions_lattice_expansion_soc}). 
Therefore, we conclude that the anomalously high resistivity with TE and SOC is caused in part by the strong phonon softening and by the high TE with SOC as shown in \cref{subfig:lattice_expansion_absolute}. 
We postulate that a better description of the TE with SOC might significantly change the computed resistivity at high temperature and that anharmonicity might be important.

\section{Conclusion}

We computed the thermodynamical properties of Pb and assessed the impact of SOC.
In all cases, we obtained relatively good agreement with experiments, especially for the TE where SOC agrees with the experimental values closely.
As for the phonon frequencies, where a strong softening is observed with SOC, the bulk modulus and its pressure derivative all undergo a similar decrease.
Only the heat capacity displayed a better agreement with experiments without SOC than with it.
To shed light on these results, we computed the electronic nesting function, the phonon linewidth $\gamma_{\vq\nu}$ and a $g$-weighted nesting function $\gamma_{\vq\nu}/\omega_{\vq\nu}$.
We showed that the nesting function might not be sufficient to explain the phonon softening, especially at $\vq=X$ where it is the largest.
We argued that it is induced by a combined effect of the EPC matrix elements being increased by the softening itself, which, in turn, would increase the softening until self-consistency is achieved.
In the end, the difference with SOC is mostly seen in the EPC matrix elements, which is stronger with SOC than without, especially at $X$, where the nesting function peaks.

We then computed the phonon-limited temperature-dependent resistivity of Pb from first principles using the state-of-the-art aiBTE formalism both with and without SOC.
We obtained results similar to previously reported resistivity data obtained via the LOVA approach, where the best agreement with experiments was obtained with SOC.
Thus, we note that the aiBTE approach does not yield a significant difference from the LOVA in the particular case of Pb.
In an attempt to improve results at high temperatures, we incorporated the TE effects by using the temperature-dependent lattice parameter obtained by minimizing the free energy.
With TE, non-linearity did appear at high temperature, but the computed resistivity is subject to a large increase with SOC as compared to that without SOC resulting in an overestimation with respect to experiments.
To understand these results, we investigated the scattering rates and noticed that TE had a similar effect than SOC, while including both makes the inverse relaxation times double, which explains the resistivity behavior with these parameters.
This non-linear effect with expanding volume correlates with the volume dependency of the $\gamma_{\vq\nu}/\omega_{\vq\nu}$ function, hinting that the strong phonon softening caused by lattice expansion transfers to the relaxation times and the resistivity. 

Among possible avenues to improve this behavior at high temperature, one could include the choice of exchange-correlation functional (XC) as it generally has strong effects on the phonon properties, especially in \ce{Pb}~\cite{grabowski_ab_2007,verstraete_density_2008}.
A comparative study of the thermodynamical properties of \ce{Pb} without SOC was already carried out by \citet{grabowski_ab_2007} where the authors compared the LDA and PBE XC.
As such, since all our results were obtained with the LDA, further investigations should compare the effects of the XC on the resistivity while considering TE and SOC.
Other XC functionals such as the recently developed r\textsuperscript{2}SCAN~\cite{furness_accurate_2020} could also be interesting to investigate, however for the latter, its implementation in DFPT frameworks has yet to be done.
Alternatively, one could investigate mechanisms beyond the QHA such as temperature dependent effective potential (TDEP)~\cite{knoop_tdep_2024} or the stochastic self-consistent harmonic approximation~\cite{monacelli_pressure_2018,bianco_second-order_2017,errea_anharmonic_2014,errea_first-principles_2013} in order to improve the overly softened phonon dispersions seen in this work.
%
Additionally, considering the phonon self-energy and including non-anharmonic contributions
could also improve the phonon dispersions.
Finally, considering a formalism beyond DFT such as the GW approximation might improve the electronic structure, especially with SOC. A better description could perhaps relieve the discrepancies of some thermodynamical properties with SOC such as the heat capacity, bulk modulus and the resistivity at high temperatures.

\begin{acknowledgments}
S. P. acknowledges support from the Fonds de la Recherche Scientifique de Belgique (FRS-FNRS) and the Walloon Region in the strategic axe FRFS-WEL-T.
This research was financially supported by the Natural Sciences and Engineering Research Council of Canada (NSERC), under the Discovery Grants program Grant No. RGPIN-2016-06666 (M.C.).
F.G. acknowledges support from the Computational Materials Sciences Program of the US Department of Energy, Office of Science, Basic Energy Sciences, under award no. DE-SC0020129.
Computations were made on the supercomputers Beluga and Narval managed by Calcul Québec and the Digital Research Alliance of Canada.
The operation of these supercomputers is funded by the Canada Foundation for Innovation, the Ministère de la Science, de l'Économie et de l'Innovation du Québec, and the Fonds de recherche du Québec – Nature et technologies (FRQ-NT).
F. A. G. acknowledges the financial support of the FRQ-NT.
\end{acknowledgments}

\bibliography{bib}
\end{document}


\section*{Supplemental Material:\\ Effects of Spin-Orbit Coupling and Thermal Expansion on the Phonon-limited Resistivity of Pb from First Principles}
    \begin{center}
        Félix Antoine Goudreault\textsuperscript{1}, Samuel Poncé\textsuperscript{2, 3}, Feliciano Giustino\textsuperscript{4, 5} and Michel Côté\textsuperscript{1}\\
    \end{center}

    \begin{itemize}
        \item[\textsuperscript{1}]Département de Physique, Université de Montréal, C. P. 6128, Succursale Centre-Ville, Montréal, Québec, H3C 3J7, Canada
        \item[\textsuperscript{2}] European Theoretical Spectroscopy Facility, Institute of Condensed Matter and Nanosciences (IMCN), Université catholique de Louvain (UCLouvain), 1348 Louvain-la-Neuve, Belgium
        \item[\textsuperscript{3}] WEL Research Institute, Avenue Pasteur 6, 1300 Wavre, Belgium
        \item[\textsuperscript{4}]Department of Physics, The University of Texas at Austin, Austin, Texas 78712, USA
        \item[\textsuperscript{5}]Oden Institute for Computational Engineering and Sciences, The University of Texas at Austin, Austin, Texas 78712, USA
        \item[\textsuperscript{6}]Department of Physics, The University of Texas at Austin, Austin, Texas 78712, USA
    \end{itemize}
    \clearpage

    \section*{Comparison of aiBTE, SERTA and LOVA}
    \begin{figure}[H]
        \centering
        \subfloat{\includegraphics{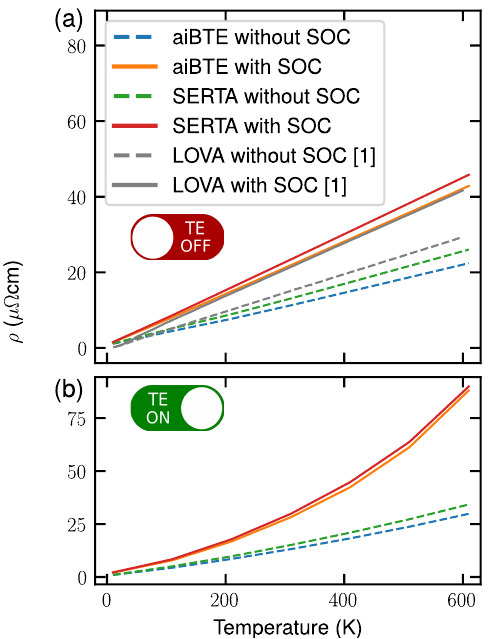}\label{suppl:ibte_vs_serta:subfigA}}
        \subfloat{\label{suppl:ibte_vs_serta:subfigB}}
        \caption{Resistivity of \ce{Pb} as a function of temperature. The solid orange, red and gray curves include SOC while the dashed blue, green and gray curves do not. The solid orange and dashed blue are computed from the full aiBTE formalism while the solid red and dashed green curves are computed within the SERTA approximation. (a) Results without TE where the lattice parameter has been fixed to the one obtained from geometry optimization. The dashed and solid gray lines are, respectively, LOVA results without and with SOC taken from Ref.~\cite{smirnov_effect_2018}. (b) Results with TE where the lattice parameter in function of temperature was extracted from the free energy minimum.}
        \label{suppl:ibte_vs_serta}
    \end{figure}
    
    \section*{Resistivity with TE and SOC at $T=\qty{610}{\kelvin}$}
    \begin{table}[H]
        \centering
        \begin{threeparttable}
            \caption{Convergence of $\rho$ with respect to the interpolated fine $\vk$ and $\vq$ grids and the energy window above and below the Fermi level.}
            \label{tab:supplemental:rho_fine_grid_conv}
            \begin{tabular}{ccc}
                \toprule\toprule
                $\vk/\vq-\text{grid}$ & Window (\unit{\electronvolt}) & $\rho$ (\unit{\micro\ohm\centi\meter})  \\
                \midrule
                 \numproduct{60x60x60} & 0.4 & 88.04\\
                 \numproduct{60x60x60} & 1.0 & 88.00\\
                 \numproduct{80x80x80} & 0.4 & 88.39\\
                 \numproduct{80x80x80} & 1.0 & 88.36\\
                 \bottomrule\bottomrule
            \end{tabular}
        \end{threeparttable}
    \end{table}

\Cref{tab:supplemental:rho_fine_grid_conv} shows the convergence behavior of the resistivity with respect to the interpolated $\vk$ and $\vq$ grids and the energy window above and below the Fermi level at $T=\qty{610}{\kelvin}$. This convergence test was conducted with SOC and using the thermally expanded lattice parameter of the corresponding temperature. 

\section*{Resistivity Coarse $\vk$ and $\vq$ Grids Convergence}

\begin{figure}[H]
    \centering
    \includegraphics{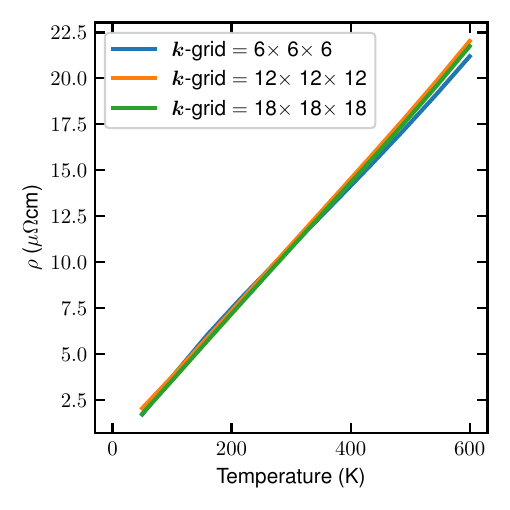}
    \caption{Resistivity of \ce{Pb} as computed from the aiBTE formalism in function of the coarse $\vk$-grid that is used as a starting point for the fine grid interpolations.}
    \label{fig:supplemental:ibte_coarse_kgrid_convergence}
\end{figure}

\Cref{fig:supplemental:ibte_coarse_kgrid_convergence} shows the behavior \ce{Pb}'s resistivity as computed from the aiBTE formalism in function of the coarse $\vk$-grid employed as basis for the fine grids interpolations. This convergence study was conducted without SOC with the lattice parameter computed using standard relaxation. Here, the energy window around the Fermi surface was set to \qty{0.2}{\electronvolt}, the coarse $\vq$ grid was fixed to \numproduct{6x6x6} and the fine $\vk$ and $\vq$ grids were fixed to \numproduct{50x50x50} points each.

\begin{figure}[H]
    \centering
    \includegraphics{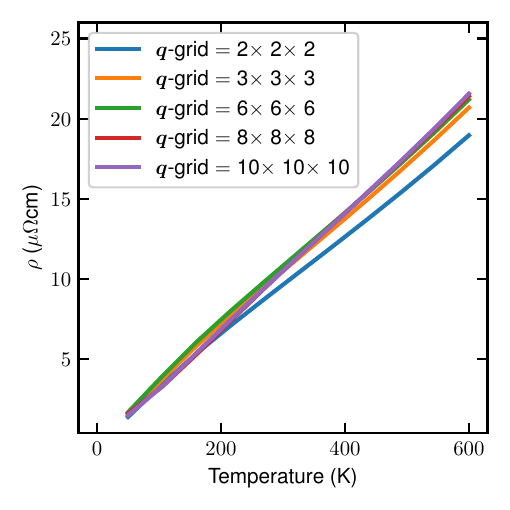}
    \caption{Resistivity of \ce{Pb} as computed from the aiBTE formalism in function of the coarse $\vq$-grid that is used as a starting point for the fine grid interpolations.}
    \label{fig:supplemental:ibte_coarse_qgrid_convergence}
\end{figure}

\Cref{fig:supplemental:ibte_coarse_qgrid_convergence} shows the behavior \ce{Pb}'s resistivity as computed from the aiBTE formalism in function of the coarse $\vq$-grid employed as basis for the fine grids interpolations. This convergence study was conducted without SOC with the lattice parameter computed using standard relaxation. Here, the energy window around the Fermi surface was set to \qty{0.2}{\electronvolt}, the coarse $\vk$ grid was fixed to \numproduct{10x10x10} and the fine $\vk$ and $\vq$ grids were fixed to \numproduct{50x50x50} points each.

\section*{Resistivity Interpolated $\vk$ and $\vq$ Grids Convergence}
\begin{figure}[H]
    \centering
    \includegraphics{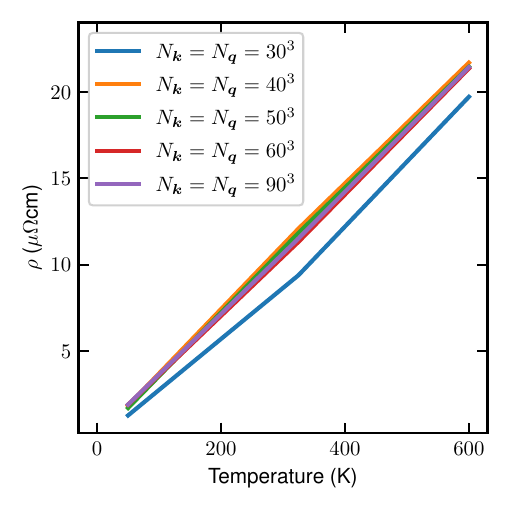}
    \caption{Resistivity of \ce{Pb} as computed from the aiBTE formalism in function of the interpolated $\vk$ and $\vq$ grids.}
    \label{fig:supplemental:ibte_fine_grid_convergence}
\end{figure}

\Cref{fig:supplemental:ibte_fine_grid_convergence} shows \ce{Pb}'s resistivity as computed from the aiBTE formalism in function of the interpolated $\vk$ and $\vq$ grid sizes (denoted respectively by $N_{\vk}$ and $N_{\vq}$). This test was conducted without SOC with the lattice parameter obtained from standard unit cell relaxation. An energy window of \qty{0.4}{\electronvolt} above and below the Fermi surface was used and a \numproduct{6x6x6} coarse $\vk$ and $\vq$ grids were employed prior to interpolation.

\section*{Resistivity Energy Window Convergence}

\begin{figure}[H]
    \centering
    \includegraphics{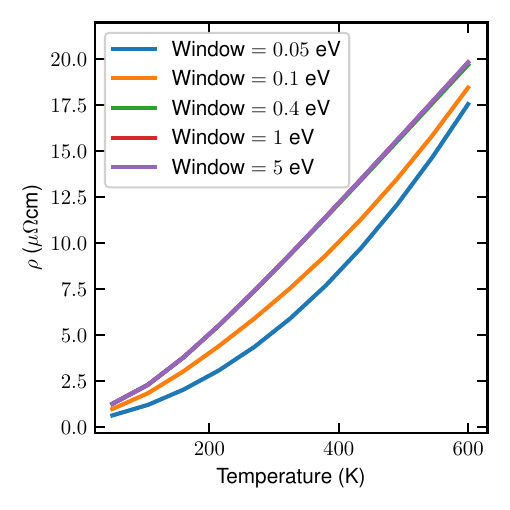}
    \caption{Resistivity of \ce{Pb} as computed from the aiBTE formalism in function of the energy window above and below the Fermi level.}
    \label{fig:supplemental:ibte_fsthick_convergence}
\end{figure}

\Cref{fig:supplemental:ibte_fsthick_convergence} shows \ce{Pb}'s resistivity as a function of the energy window above and below the Fermi level. This test was conducted with a \numproduct{8x8x8} coarse $\vk$-grid and a \numproduct{6x6x6} coarse $\vq$-grid which were interpolated on finer \numproduct{30x30x30} grids.

\section*{Phonon frequencies at $\vq=X$}
\begin{figure}[H]
    \centering
    \includegraphics{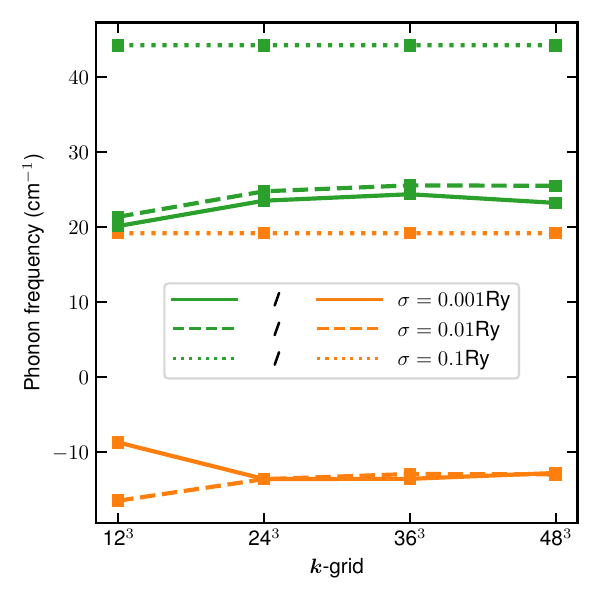}
    \caption{\label{suppl:phonon_smearing_convergence_high_volume}Phonon frequencies at $\vq=X$ in function of the $\vk$-grid sampling of the first Brillouin Zone for different electronic smearings ($\sigma$) where the primitive cell volume has been expanded by \qty[retain-explicit-plus]{+12}{\percent} with respect to the relaxed volume $\Vdft$. Orange lines represent the frequencies for the transverse modes ($T_1$ and $T_2$) while the green lines represent frequencies for the longitudinal mode ($L$). Here, a Gaussian smearing function as been used.}
\end{figure}

\begin{figure}[H]
    \centering
    \includegraphics{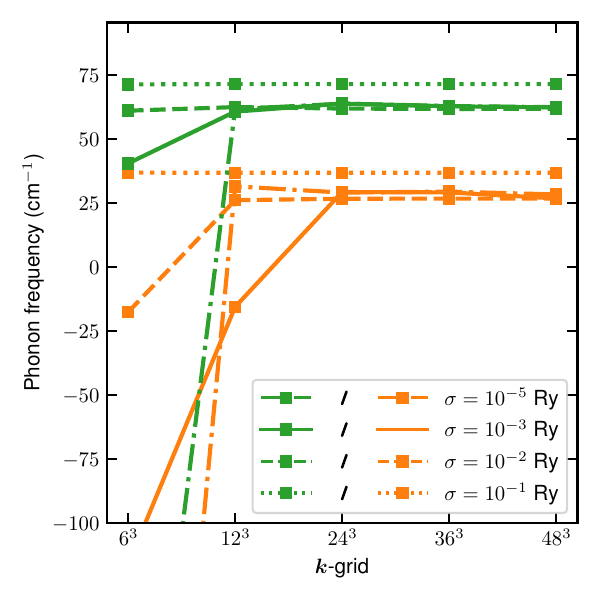}
    \caption{\label{suppl:phonon_smearing_convergence_relax_volume}Phonon frequencies at $\vq=X$ in function of the $\vk$-grid sampling of the first Brillouin Zone for different electronic smearings ($\sigma$) where the primitive cell volume is $\Vdft$. Orange lines represent the frequencies for the transverse modes ($T_1$ and $T_2$) while the green lines represent frequencies for the longitudinal mode ($L$). Here, a Gaussian smearing function as been used.}
\end{figure}

\begin{figure}[H]
    \centering
    \includegraphics[width=0.95\linewidth]{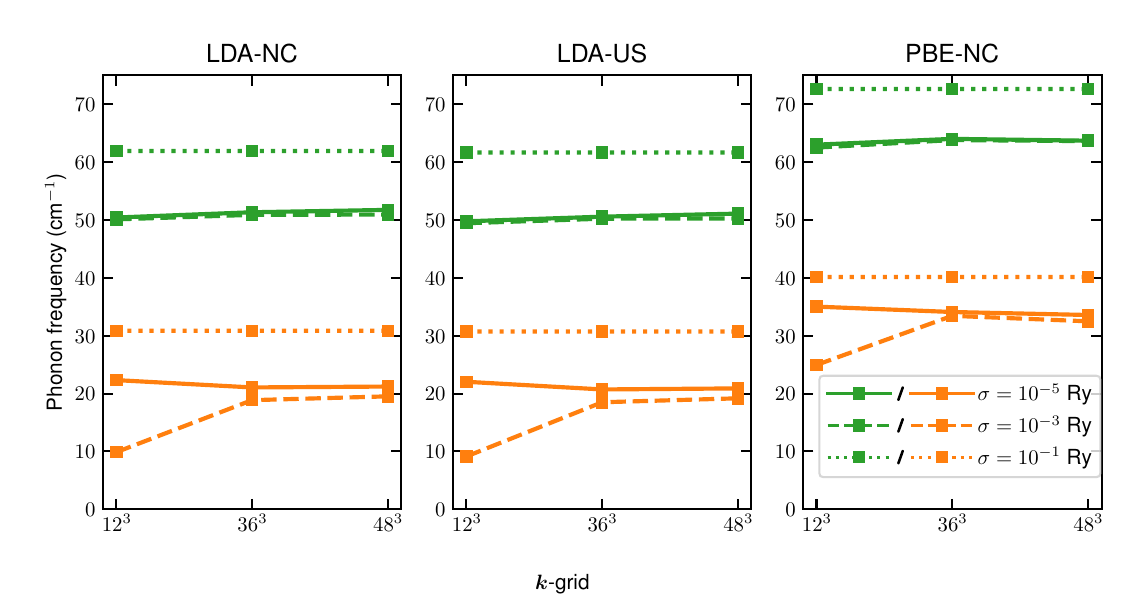}
    \caption{\label{suppl:phonon_smearing_convergence_xc}Phonon frequencies at $\vq=X$ in function of the $\vk$-grid sampling of the first Brillouin Zone for different electronic smearings. The frequencies were computed at the experimental lattice parameter at room temperature $a^{\text{exp}}=\qty{4.95}{\angstrom}$ taken from Ref.~\cite{zagorac_recent_2019}. The left and middle panel make use of the LDA exchange-correlation functional where the left panel employs a norm-conserving pseudopotential while the middle one employs the ultra soft pseudopotential which was used for all calculations of the main manuscript. The right panel employs a PBE functional with a norm-conserving pseudopotential. The norm-conserving pseudopotentials were obtained from the Pseudodojo project~\cite{van_setten_pseudodojo:_2018}.}
\end{figure}

\bibliographystyle{plain}
\bibliography{bib}